\documentclass[aps,epsfig,graphicsfloatfix,mathbbm,a4paper,nofootinbib,twocolumn]{article}

\usepackage{amsmath,amsfonts,amssymb,graphics,graphicx,epsfig,color,times,bbm,titletoc}

\titlecontents{subparagraph}[5.25em]{}{\contentslabel{4.6em}}{}{\titlerule*[0.6pc]{.}\contentspage}

\begin{document}

\bibliographystyle{apsrev}
\newcommand{\spec}{\rm{spec}}
\newcommand{\id}{\rm{id}}
\newcommand{\R}{\mathbbm{R}}
\newcommand{\cB}{{\cal B}}\newcommand{\cS}{{\cal S}}
\newcommand{\CP}{{\mathbb{C}}{\bf P}}
\newcommand{\cH}{{\cal H}}
\newcommand{\rr}{\mathbbm{R}}
\newcommand{\cT}{{\cal T}}
\newcommand{\cV}{{\cal V}}
\newcommand{\cL}{{\cal L}}
\newcommand{\E}{{\cal E}}
\newcommand{\cX}{{\cal X}}
\newcommand{\V}{{\cal V}}
\newcommand{\cc}{{\cal{C}}}
\newcommand{\ii}{\mathbbm{1}}
\newcommand{\cM}{{\mathcal M}}
\newcommand{\ra}{\rightarrow}
\newcommand{\C}{\mathbb{C}}
\newcommand{\1}{\mathbbm{1}}
\newcommand{\F}{\mathbbm{F}}
\newcommand{\h}{\frak{H}}
\newcommand{\tr}[1]{{\rm tr}\left[#1\right]}
\newcommand{\gr}[1]{\boldsymbol{#1}}
\def\>{{\rangle}}
\def\<{{\langle}}
\newcommand{\be}{\begin{equation}}
\newcommand{\ee}{\end{equation}}
\newcommand{\bea}{\begin{eqnarray}}
\newcommand{\eea}{\end{eqnarray}}
\newcommand{\ket}[1]{|#1\rangle}
\newcommand{\bra}[1]{\langle#1|}
\newcommand{\avr}[1]{\langle#1\rangle}
\newcommand{\red}[1]{{\bf \textcolor{red}{{#1}}}}
\newcommand{\D}{{\cal D}}
\newcommand{\eq}[1]{Eq.~(\ref{#1})}
\newcommand{\ineq}[1]{Ineq.~(\ref{#1})}
\newcommand{\sirsection}[1]{\section{\large \sf \textbf{#1}}}
\newcommand{\sirsubsection}[1]{\subsection{\normalsize \sf \textbf{#1}}}
\newcommand{\ack}{\subsection*{\normalsize \sf \textbf{Acknowledgements}}}
\newcommand{\front}[5]{\title{\sf \textbf{\Large #1}}
\author{#2 \vspace*{.4cm}\\
\footnotesize #3}
\date{\footnotesize \sf \begin{quote}
\hspace*{.2cm}#4 \end{quote} #5} \maketitle}
\newcommand{\eg}{\emph{e.g.}~}

\newcommand{\proofend}{\hfill\fbox\\\medskip }

\newcommand{\As}{\mathfrak{S}(\Ao)} 
\newcommand{\ABs}{\mathfrak{S}(\Ao,\Bo,\ldots)} 
\newcommand{\Ascomp}{\As^\perp} 
\newcommand{\Bs}{\mathcal{B}} 
\newcommand{\Bscomp}{\mathcal{B}^\perp} 
\newcommand{\os}{\mathfrak{S}} 

\newcommand{\rank}[1]{\mathrm{rank}(#1)} 


\newtheorem{theorem}{Theorem}
\newtheorem{proposition}{Proposition}

\newtheorem{lemma}{Lemma}

\newtheorem{example}{Example}
\newtheorem{corollary}{Corollary}

\newcommand{\proof}[1]{{\it Proof.} #1 $\proofend$}

\title{\begin{center}
    {  Are problems in Quantum Information Theory (un)decidable?}
\end{center}}
\author{Michael M. Wolf\thanks{Department of Mathematics, Technische Universit\"at M\"unchen, Email:  \rm{m.wolf@tum.de}}, Toby S. Cubitt\thanks{Departamento de Analisis Matematico, U. Complutense de Madrid}, David Perez-Garcia\thanks{Departamento de Analisis Matematico, IMI, U. Complutense de Madrid}}
\date{\today}
\maketitle
\begin{abstract} This note is intended to foster a discussion about the extent to which typical problems arising in quantum information theory are algorithmically decidable (in principle rather than in practice). Various problems in the context of entanglement theory and quantum channels turn out to be decidable via quantifier elimination as long as they admit a compact formulation without quantification over integers. For many asymptotically defined properties which have to hold for all or for one $n\in\mathbb{N}$, however, effective procedures seem to be difficult  if not impossible to find. We review some of the main tools for (dis)proving decidability and apply them to problems in quantum information theory. We find that questions like "can we overcome fidelity 1/2 w.r.t. a two-qubit singlet state?" easily become undecidable. A closer look at such questions might rule out some of the ``single-letter'' formulas sought in quantum information theory.
\end{abstract}

\tableofcontents

\section{Introduction}
The elementary objects of quantum information theory are often rather simple small dimensional matrices---for instance representing density matrices, quantum channels or observables. In spite of their innocent looking mathematical description, their properties can be difficult to compute quantitatively or even to decide qualitatively: Is a given quantum state entangled? Is it distillable? Does it admit a local hidden variable description? Is a quantum channel a mixture of unitaries? Does it have non-zero quantum capacity? None of these questions seems to have a simple answer.

From an abstract point of view, the difficulty in deciding such questions arises from quantification over infinite sets. That is, answering these questions requires taking for instance all decompositions, protocols, models, codings, etc. into account.  Moreover, in particular information theoretic questions often involve asymptotic limits regarding the numbers of copies, rounds within a protocol, measurements,  etc..

In the present work we analyse problems like the ones mentioned above from the point of view of algorithmic decidability. This means we don't ask (at least not on the following pages) whether there are resource-efficient algorithms, we rather want to know whether there is \emph{any effective algorithm}, irrespective of the scaling of resources. 

One of our motivations---although the following work is only a minor step in this direction---is to address the (im)possibility of ``single-letter'' formulas/criteria, especially for asymptotically defined quantities. Clearly, those arise not only in information theoretic contexts but rather naturally also in statistical physics, where phases and their properties are defined in the thermodynamic limit. In the present paper we will, however, focus exclusively on problems occurring in quantum information theory.\footnote{Undecidablility  of properties of  quantum spin chains will be the content of a separate article \cite{CPW12}.} Moreover, we will restrict ourselves to decidability problems  and, for the moment, leave  closely related questions of  (quantitative) computability aside.\footnote{Note that a quantity, e.g. some capacity, might be computable although it is undecidable whether or not it is different from zero.}

We will go through three stages: decidable problems (Sec.\ref{sec:decidable}), undecided problems (Sec.\ref{sec:undecided}), and undecidable problems (Sec.\ref{sec:undecidable}). The notion of decidability which we have in mind is the one of recursive function theory rather than, for instance, the one of real computation (as in \cite{BCSS97}).

Others have followed similar lines: decidability problems in quantum automata were for instance studied in \cite{BJKP05,DJK05,Hir08} and results on quantum implications and interpretations of the undecidable 'matrix mortality' problem are found in \cite{EMG11}.

\section{Decidable problems}\label{sec:decidable}
As already mentioned in the introduction, deciding properties of quantum information theory often involves taking for instance all decompositions, protocols, models, codings, etc. into account. Mathematically this can be expressed in terms of quantification over the respective sets. In this section we discuss a general tool for proving decidability via elimination of all these quantifiers.

\subsection{A general tool from Tarski and Seidenberg}
Let us introduce the general idea via a simple example: if $a_1$ and $a_2$ are real numbers, then the equation $a_1x^2>a_2$ has a real solution iff $a_2$ is negative or $a_1$ is positive. More formally,
\bea\label{eq:TS_exp1} &\exists x& a_1x^2>a_2,\\ \label{eq:TS_exp2}&\Leftrightarrow & a_2<0 \vee a_1>0.\eea
That is, in Eq.(\ref{eq:TS_exp2}) we have a quantifier-free expression which has the same truth value as the initial formula in Eq.(\ref{eq:TS_exp1}).

Tarski \cite{Tar51} and Seidenberg \cite{Sei54} have shown that such an elimination of  both existential $\exists$ and universal $\forall$ quantifiers is possible in a more general context. Sufficient requirements are that the initial formula  (i) contains real variables and parameters together with the objects and relations $1,0,+,\cdot,>,=$ of the ordered real  field, and  (ii) is expressed in first-order logic. The latter means that in addition to quantifiers over individual variables the formula is  allowed to involve Boolean operations  $\vee, \wedge,\neg$ and brackets for unambiguous readability. Something like $\forall n\in\mathbb{N} \ldots$ or quantification over sets, however, is not allowed. A more precise formulation of the underlying theorem is (cf.\cite{Mar08}):
\begin{theorem}[Tarski-Seidenberg, quantifier elimination]\label{thm:TSQE}
Let $R$ be a real closed field. Given a finite set $\{f_i(x,a)\}_{i=1}^k$ of polynomial equalities and inequalities with variables $(x,a)\in R^n\times R^m$ and  coefficients in $\mathbb{Q}$. Let $\phi(x,a)$ be a Boolean combination of  the $f_i's$ (using $\vee, \wedge,\neg$) and
\be\Psi(a):= \Big(Q_1 x_1\ldots Q_n x_n: \phi(x,a)\Big), \ \ Q_j\in\{\exists,\forall\}.\ee
Then there exists a formula $\psi(a)$ which is (i) a quantifier-free Boolean combination of finitely many polynomial (in-) equalities with rational coefficients, and (ii) equivalent in the sense
\be \forall a: \ \ \big(\psi(a) \Leftrightarrow \Psi(a)\big).\ee
Moreover, there exists an effective algorithm which constructs the quantifier-free equivalent $\psi$ of any such formula $\Psi$.
\end{theorem}
A \emph{real closed field} is an ordered field which satisfies the intermediate value theorem for polynomials. Examples are the field $\R$ of real numbers, the field of computable real numbers and the field $\R_A$ of algebraic real numbers.
Algorithms for quantifier elimination are effective but not generally efficient \cite{BPR94}. For a comparison of some of the implemented algorithms see \cite{DSW98}.

Suppose now we want to decide whether or not some $\Psi(a)$ is true for given $a$. Then we can use quantifier elimination in order to obtain a simpler, quantifier-free expression $\psi(a)$, so that it remains to check a set of (in-)equalities for polynomials in the $a_i$'s, like in Eq.(\ref{eq:TS_exp2}) of our example. In order for this to be feasible we need a \emph{computable ordered field} in which the basic operations $\cdot,+$ as well as the relations $=,>$ are computable. The latter is not the case for the fields of real or computable real numbers---being able to compute an arbitrary number of decimals for $a_1$ is not sufficient for an effective procedure for deciding $a_1>0$.

In the following we restrict the parameters $a$ to any computable ordered field $\R_c$ which is real closed and contains the field of algebraic numbers $\R_{alg}$. The latter forms a computable ordered and real closed field on its own \cite{LM70} so that we may simply set $\R_c=\R_{alg}$. With some care, however, including (computable) transcendental numbers is possible: one can for instance start with non-algebraic extension fields of $\mathbb{Q}$ as in \cite{Lev09} and then use the fact that any real closure of a computable ordered field is again a computable ordered field \cite{Mad70}. By $\C_c:=\R_c(\sqrt{-1})$ we will denote the algebraic closure of $\R_c$. If for instance $\R_c=\R_{alg}$, then $\C_c$ is the field of complex algebraic numbers.

In spite of all this, we will apply Thm.\ref{thm:TSQE} to $R=\R$, since the problems we will discuss involve quantification over the reals.

Problems arising in quantum information theory often involve quantification over matrices or linear maps, such as unitaries, positive semidefinite matrices, positive or completely positive maps, matrices with bounded rank or bounded norm, etc.
Moreover, relevant inequalities are often w.r.t. the semidefinite partial ordering of Hermitian matrices.

Before discussing specific problems from quantum information theory, we will convince ourselves that the Tarski-Seidenberg trick works as well if quantification is over any of the sets just mentioned. We will denote by $\cM_d(G)$ the set of $d\times d$ matrices with entries in $G$.
\begin{lemma}\label{lem:polysets} Let $x\in\R^{2d^2}$ be a vector containing the real and imaginary parts of a matrix $X\in\cM_d(\C)$. The membership relation $X\in S$ is equivalent to a Boolean combination of polynomial (in-)equalities in $x$ with integral coefficients, if  $S\subset\cM_d(\C)$ is one of the following sets:
\begin{enumerate}
\item the set \emph{positive semidefinite} matrices,
\item the set of \emph{density matrices} ,
\item the set of \emph{unitaries},
\item the set of Hermitian matrices with \emph{bounded norm} $||X||_p\leq 1$ for any $p\in \mathbb{N}\cup\{\infty\}$,
\item the set of matrices with \emph{constrained rank} $\rm{rank}(X)=r$ for any $r\leq d$.
\end{enumerate}
\end{lemma}
\proof{1. follows from the fact that $X\geq 0$ holds iff $X=X^\dagger$ and all the principal minors of $X$ are non-negative.

2. follows from 1. and the linear equation $\tr{X}=1$.

3. follows from the quadratic equation $X^\dagger X=\1$.

4. since $||X||_\infty\leq 1$ iff $(\1-X^\dagger X)\geq 0$ we can use 1. For $p\in\mathbb{N}$ we obtain a polynomial inequality from $\tr{X^p}\leq 1$.

5. follows from the fact that the rank is the size of the largest non-vanishing minor.
}

In general, sets which can be characterized by means of polynomial (in)-equalities in a finite-dimensional real vector space are called \emph{semialgebraic}.

We will call a relation $F$ on $\cM_{d_1}(\C)\times\ldots\times\cM_{d_K}(\C)$ a \emph{polynomial matrix equation with rational coefficients} iff it can be expressed as $\hat{F}(M_1,\ldots,M_K)\rhd 0$, where $\hat{F}$ is a matrix (possibly one-dimensional) whose entries are polynomials in the entries of the $M_i$'s with rational coefficients and $\rhd\in\{>,\geq,=\}$. The inequalities refer to the partial order induced by the cone of positive semidefinite matrices.

The above Lemma together with the Tarski-Seidenberg theorem now leads to the following useful observation:
\begin{corollary}\label{Cor:TSQE}
Let $A$ be a tuple of matrices with entries in $\C_c$ and let $\{F_i(X,A)\}_{i=1}^k$ be a finite set of polynomial matrix equations with rational coefficients and variables $X\in\cM_{d_1}(\C)\times\ldots\times\cM_{d_n}(\C)$. Given a Boolean combination $\phi(X,A)$ of the $F_i$'s and
\be\Psi(A):= \Big(Q_1 (X_1\in S_1)\ldots Q_n (X_n\in S_n): \phi(X,A)\Big), \ee
where $Q_j\in \{ \exists,\forall \}$ and the $S_i$'s are semialgebraic sets (e.g. those specified in Lemma \ref{lem:polysets}). Then there exists an effective algorithm which decides $\Psi(A)$.
\end{corollary}
This corollary uses Thm.\ref{thm:TSQE} after we have expressed everything in terms of the real and imaginary parts of the matrix entries, used Lemma \ref{lem:polysets} and brought the equation into prenex normal form. After the quantifier elimination we then exploit that $A$ is specified over a computable ordered field which finally admits an effective algorithm.

Let us apply this observation to some standard problems in the context of entanglement and quantum channels. All the problems discussed in the remainder of this section are such that they involve quantification ($\exists,\forall$) over semialgebraic sets, like the ones in Lemma \ref{lem:polysets}, and a semialgebraic equation to be checked. Consequently, Cor.\ref{Cor:TSQE} applies to all of them.
\subsection{Entanglement}\label{sub:ent}
A density matrix $\rho\in \cM_{d}(\C)^{\otimes n}$ is called \emph{classically correlated} or \emph{unentangled} \cite{Wer89} iff there exist sets of density matrices $\{\rho_i^{(\alpha)}\in\cM_{d}(\C)\}_{i=1}^{d^{2n}}$ for $\alpha=1,\ldots,n$  and a probability vector $\lambda\in\R^{d^{2n}}$ such that\footnote{The bound on the range of $i$, which is not part of the definition, follows from Caratheodory's theorem.}
\be \rho=\sum_i \lambda_i \rho_i^{(1)}\otimes\cdots\otimes\rho_i^{(n)}.\ee
If the entries of $\rho$ are taken from $\C_c$, then the question whether or not $\rho$ is entangled is of the form in Cor.\ref{Cor:TSQE} and can thus be effectively decided.
\subsection{$n$-distillability}
A density matrix $\rho\in\cM_d(\C_c)^{\otimes 2}$ is $n-$\emph{distillable} iff there exists a rank-2 matrix $Y\in\cM_{d^n}(\C)$ such that
\be
\<y|\big(\rho^{\otimes n}\big)^{T_1}|y\> <0,
\ee
where $|y\>=\sum_{k,l}Y_{k,l}|k,l\>$ and $ ^{T_1}$ denotes the partial transposition (both understood w.r.t. the bipartite rather than the $n$-partite partitioning). For fixed $n\in\mathbb{N}$ this is again such that Cor.\ref{Cor:TSQE} implies an effective decision procedure.

\subsection{Existence of an $(n,m)$-hidden variable model}\label{sub:LHV}
Let $P(i,j|k,l)$ be a conditional probability distribution so that $\sum_{i,j=1}^m P(i,j|k,l)=1$ for all $k,l=1,\ldots,n$. $P\in\R^{m^2n^2}$ admits a \emph{local hidden variable} (LHV) model \cite{WW01b} iff there exists a non-negative matrix $\Lambda\in\cM_{m^n}(\R)$ satisfying $\sum_{a,b}\Lambda_{a,b}=1$ such that for all $i,j,k,l$ it holds that
\be P(i,j|k,l)=\sum_{a,b}\Lambda_{a,b}\;\delta_{a(k),i}\;\delta_{b(l),j},\ee where $a,b\in\{1,\ldots,m\}^n$.
We say that a bipartite density matrix $\rho\in\cM_d(\C_c)^{\otimes 2}$ admits an $(n,m)$-hidden variable description iff the distribution
\be P(i,j|k,l):=\tr{\rho\big(Q_i^{(k)}\otimes P_j^{(l)}\big)}\label{eq:PrhoQP}\ee
admits a LHV model for all sets of POVMs $\{Q_i^{(k)}\in\cM_d(\C)\}$ and $\{P_j^{(l)}\in\cM_d(\C)\}$, where $i,j$ label the $m$ effect operators of each of the $2n$ POVMs. Since the latter are again characterized by semialgebraic means, the question whether or not a quantum state admits an $(n,m)$-hidden variable description is effectively decidable by Cor.\ref{Cor:TSQE}.  Note that the quantifier structure of this problem involves also universal quantifiers as it is of the form $\forall Q \forall P \exists \Lambda \ldots$.

\subsection{Existence of a $d-$dimensional quantum representation}\label{sub:qrep}

Conversely, we may ask whether a probability distribution $P\in\R_c^{m^2n^2}$ admits a $d-$dimensional quantum representation in the sense that there exists a density matrix $\rho\in\cM_d(\C)^{\otimes 2}$ and sets of POVMs $\{Q_i^{(k)}\in\cM_d(\C)\}$ and $\{P_j^{(l)}\in\cM_d(\C)\}$ such that Eq.(\ref{eq:PrhoQP}) holds for all $i,j,k,l$.  Again Cor.\ref{Cor:TSQE} applies.

\subsection{Birkhoff property}\label{sub:Birkhoff}
Let us turn to quantum channels $T:\cM_d(\C)\ra\cM_d(\C)$, i.e., completely positive, trace-preserving linear maps. For a unital quantum channel, i.e., one which satisfies $T(\1)=\1$ in addition, there may exist an $n\in\mathbb{N}$, unitaries $\{U_i\in\cM_{d^n}(\C)\}$ and a probability vector $\lambda$ such that
\be T^{\otimes n}(\rho)= \sum_i\lambda_i U_i\rho U_i^\dagger\quad\text{for all } \rho\label{eq:Birkhoff}.\ee
For $d=2$ this holds in fact for all unital quantum channels (even for $n=1$). However, it fails to be generally true if $d\geq 3$ \cite{HM11}. Since we can bound the range of the index by Caratheodory's convex hull theorem to  $i=\{1,\ldots, d^{4n}\}$ the problem of deciding whether a unital  quantum channel can be represented as in Eq.(\ref{eq:Birkhoff}) again falls into the field of activity of Cor.\ref{Cor:TSQE} if $n$ is fixed and the channel's Choi matrix or its Kraus operators are over $\C_c$.

\subsection{$n-$shot zero-error capacity}\label{sub:zeroerror}

A quantum channel $T:\cM_d(\C)\ra\cM_d(\C)$ has a non-vanishing $n-$\emph{shot classical zero-error capacity}, iff there exist density matrices $\rho,\sigma\in\cM_{d^n}(\C)$ such that
\be \tr{T^{\otimes n}(\rho)T^{\otimes n}(\sigma)}=0.\ee
One more time, since the quantification is over the set of density matrices and all equations are (semi-)algebraic, Cor.\ref{Cor:TSQE} applies for any fixed $n\in\mathbb{N}$.

More generally, the $n-$shot classical zero-error capacity is greater than $\log m/n$ iff there exist $m$ density matrices $\rho_i\in\cM_{d^n}(\C)$ ($i=1\dots m$) such that
\be \tr{T^{\otimes n}(\rho_i)T^{\otimes n}(\rho_j)}=0\ee
for all $i\neq j$ and Cor.\ref{Cor:TSQE} applies once again for any fixed $n,m\in\mathbb{N}$.

\subsection{Additive minimal output entropy}\label{sub:additivity}

For any $p\in\mathbb{N}\cup\{\infty\}$ define a functional $\nu_p$ on the set of quantum channels by
\be \nu_p(T):=\sup\big\{||T(\rho)||_p\big|\rho\geq 0\wedge\tr{\rho}=1\big\},\ee
and consider the decision problem whether or not
\be \nu_p\big(T\otimes T'\big)=\nu_p(T)\nu_p(T'),\label{eq:mult1}\ee
holds for all quantum channels $T':\cM_d(\C)\ra\cM_d(\C)$ of a fixed dimension.
Since the l.h.s. in Eq.(\ref{eq:mult1}) is always lower bounded by the r.h.s., this problem can be phrased using quantifiers (over the respective sets) in the following way:
\be \exists\rho_1\forall{T'} \exists\rho_2\forall\rho_{12}:||(T\otimes T')(\rho_{12})||_p \leq||T(\rho_1)||_p||T'(\rho_2)||_p.\label{eq:addquant}\ee
Note that all sets and equations are semialgebraic. Hence, for fixed $p$ and $d$, Cor.\ref{Cor:TSQE} implies the existence of an effective algorithm which, upon input of a finite-dimensional channel $T$ which is defined over $\C_c$, determines the truth-value of Eq.(\ref{eq:addquant}).

\section{Undecided problems}\label{sec:undecided}
Consider the problems discussed in \ref{sub:ent} to~\ref{sub:additivity} of the last section again.  Apart from \ref{sub:ent}, all problems share the appearance of one or more integers whose quantification leads to a more fundamental problem. However, adding such a ``$\exists n\in\mathbb{N}$'' or ``$\forall n\in\mathbb{N}$'' disables the Tarski-Seidenberg tool and leaves us  with problems for which, to the best of our knowledge, no effective procedure is currently known. The catch is, that for every $n\in\mathbb{N}$ we have an effective procedure, but as long as there is no upper bound on the ``largest relevant'' $n$, these cannot be combined to a universal effective algorithm. Of course, for some of the discussed problems quantification over the remaining integer(s) may eventually again lead to a semialgebraic set for which membership can be decided using Tarski-Seidenberg.

Let us have a closer look, problem by problem.
\subsection{Distillability:} A bipartite quantum state is called \emph{distillable} iff there exists an $n\in\mathbb{N}$ such that it is $n-$distillable. In this case, it was shown in \cite{Wat04} that there is indeed no dimension-dependent upper bound: for any $n$ there exist a density matrix acting on $\C^9\otimes\C^9$ such that the state is distillable but not $n-$distillable. At present it is not known whether or not being not distillable coincides with the property of the density matrix having a positive partial transpose. If this would be true, then distillability would be decidable since, as we've seen in Lemma\ref{lem:polysets}, positive semi-definiteness is a semialgebraic relation. Conversely, this means that undecidability of distillability would imply the existence of undistillable states whose partial transpose is not positive---with all its puzzling consequences \cite{SST00,VW02b}.\vspace*{5pt}

\subsection{LHV models:} Regarding the existence of a universal hidden variable model (one which holds for all $n,m\in\mathbb{N}$) for a given density matrix, not much seems to be known. There are quantum states which are entangled and nevertheless admit such a universal LHV model \cite{Wer89,Bar02}. However, it is also clear that there are states which admit a LHV description for $n=m=2$ but fail to do so for larger $n$ or $m$.\vspace*{5pt}

\subsection{Quantum representation for correlations:} The relevant integral parameter in the problem raised in \ref{sub:qrep} is $d\in\mathbb{N}$, so that the underlying question is: given a probability distribution $P(i,j|k,l)$ is there any $d\in\mathbb{N}$ for which there exists a quantum representation. To the best of our knowledge, this question is still open. There is some evidence \cite{PV10} that already for $ n=3, m=2$ the dimension  $d$ might be generally unbounded or even infinite. The recent connection between Tsirelson's problem and the long standing Connes' embedding problem \cite{JNPPSW11,Fri10} also points in this direction. This could explain why  attempts to find a characterization of the set of quantum correlations, like the hierarchy of semidefinite relaxations given in \cite{NPA07}, lead in the worst case scenario to infinite running times. \vspace*{5pt}

\subsection{Birkhoff property:} For $d=3$ there are unital quantum channels known for which Eq.(\ref{eq:Birkhoff}) does not hold for $n=1$ but becomes true for $n=2$ \cite{MW09}. It is also known that the set of channels for which there exists such an $n\in\mathbb{N}$ is a subset of the so-called \emph{factorizable} channels \cite{HM11}.

\subsection{Zero error-capacities:}
Already in the case of the zero-error capacity of \emph{classical} channels, it is known that there is no upper bound on the block-length $n$ required to achive the capacity (see \cite{KO98} and references therein). Since classical channels are a special case of quantum channels, this carries over immediately to the classical zero-error capacity of quantum channels.

In the important case of deciding whether the zero-error capacity vanishes, classical channels are no longer any use, since it is trivial to see that the $n-$shot zero-error capacity of a classical channel vanishes iff it vanishes for $n=1$. Thanks to entanglement, this is \emph{not} true for quantum channels. A simple implication of the superactivation results of \cite{CCH09,Run09} is that there exist quantum channels for which the $1-$shot capacity vanishes but the $2-$shot capacity does not,
but it is not known at present whether this extends to arbitrary $n$.

\subsection{(Non-)additivity of other capacities:}
The maximal output $p-$norm is known not to be multiplicative in general \cite{HW02,WH08,Has09}. Channels for which multiplicativity of the form $\nu_p(T^{\otimes 2})=\nu_p(T)^2$  has been proven, are often such that $\nu_p(T\otimes T')=\nu_p(T)\nu_p(T')$ holds for any $T'$ irrespective of its dimension.

The failure of additivity of various single-shot quantities (such as the minimal output entropy corresponding to $\nu_p$ for $p\ra 1$) leads to the fact that we do currently neither have a single-letter formula for the classical capacity of quantum channels nor for its quantum capacity. The latter is known to be itself non-additive in the strong sense that $0+0>0$ \cite{SY08}. Questions like ``is the quantum capacity zero?'' or, if it is, ``can it be activated by another zero-capacity channel'' also seem undecided.

\subsection{More problems}
The above list can easily be extended as there are numerous problems which involve quantification over one or several integers. Further examples would be: is a quantum channel implementable by means of LOCC (for which there is no a priori bound on the number of required rounds)? Can a source or a channel asymptotically simulate another one at a pre-given rate and under given constraints? Etc.

\section{Undecidable problems}\label{sec:undecidable}
\subsection{Tools for proving undecidability}
Before we briefly review some tools for proving that a problem is \emph{algorithmically undecidable}, let us recall what this means and implies. The reader familiar with the basic notions of computability theory may skip this subsection. The reader not satisfied by it may consult \cite{Cut80,Bar77}.

Informally, an \emph{algorithm} is an effective procedure based on finitely many instructions which can be carried out in a stepwise fashion. Crucial for this notion is that there has to be a finite description. Any formal definition has to specify \emph{who} is carrying out \emph{what kind of} instructions. The two historically first frameworks which made this precise were \emph{Turing machines} and \emph{recursive function theory}. However, any  other reasonable framework people came up with turned out to be no more powerful than those. That this will continue to hold in the future is the content of the \emph{Church-Turing thesis}.

  For a decision problem to be algorithmically undecidable, an infinite domain is necessary. Suppose there were only finitely many, say $N$, distinct inputs for which we want a yes-no question to be decided. Then we may write one algorithm for each of the $2^N$ truth-value assignments and since this exhausts all possibilities one of the algorithms will return the correct answer for all the inputs.\footnote{We tacitly assume the \emph{tertium non datur}.} Finite problems are thus decidable which, for the problems we have discussed, implies that  introducing  ``$\epsilon$-balls'' will make them algorithmically decidable (although we still wouldn't know the algorithm). An infinite domain, on the other hand, should be countable infinite, since otherwise the inputs have no finite description.

Requiring algorithms to have a finite description and considering decision problems with countable infinite domains already implies that there are undecidable problems (viewed as subsets of $\mathbb{N}$) since algorithms are countable but decision problems aren't, i.e., $|\mathbb{N}|<|2^\mathbb{N}|$.

\subparagraph{The halting problem} is the mother of all undecidable problems. This is not only true historically, but also technically: most problems which are known to be undecidable are proven to be so by showing that the halting problem can (directly or indirectly) be reduced to them. The halting problem asks whether a Turing machine halts upon a given input. If we identify the inputs with natural numbers, then the subset of $\mathbb{N}$ upon which a fixed Turing machine halts is called \emph{recursively enumerable}---a notion we will come back to later. Clearly, membership in a recursively enumerable set cannot generally be decidable since otherwise the halting problem would be decidable.

\subparagraph{Post's correspondence problem } (PCP) is an example of an undecidable problem to which the halting problem can be reduced. It is frequently used as an intermediate step in undecidability proofs by reducing it in turn to the problem under consideration.

Let $A$ be a finite alphabet for which we consider the set of words $A^*$  as a monoid with respect to which $X,Y:K\ra A^*$  are two homomorphisms from $K=\{1,\ldots,k\}$. Post`s correspondence problem is then to decide whether or not there is a non-empty word $w\in A^*$ such that $X(w)=Y(w)$. An equivalent, possibly more intuitive, depiction of the problem is to think about $k$ types of dominos and ask whether they admit a finite chain in which the upper row coincides with the lower row.

While PCP is decidable if $k=2$, it's known to be undecidable for $k\geq 7$ \cite{MS05}. In fact, it is undecidable whether there is a so-called \emph{Claus instance} solution of the form  $1w7$ with $w\in\{2,\ldots,6\}^*$ (cf. \cite{Har09}).

The main reason why we emphasize PCP, is that it can be reduced to problems involving products of matrices---something we will make use of in the next subsection. The main tool in this context is a monoid morphism due to Paterson \cite{Pat70}: take $A=\{1,\ldots,m\}$ and define an injection $\sigma:A^*\ra\mathbb{N}$ as the $m-$adic representation of words, i.e., $\sigma(w)=\sum_{j=1}^{|w|} w_j m^{|w|-j}$ where $|w|$ denotes the length of the word and $\sigma$ maps the empty word onto $0$. Note that $\sigma(uw)=m^{|w|}\sigma(u)+\sigma(w)$. The map $\gamma:A^*\times A^*\ra\cM_3(\mathbb{N}_0)$
\be
\gamma(u,w):=\left(\begin{array}{ccc}
m^{|u|} & 0 & 0 \\
0 & m^{|w|} & 0 \\
\sigma(u) & \sigma(w) & 1
\end{array} \right),
\ee is then a monoid monomorphism, i.e., it holds that $\gamma(uu',ww')=\gamma(u,w)\gamma(u',w')$. Defining vectors $|x\>:=(1,-1,0)^T$ and $\<y|:=(0,0,1)$ this implies that $u=w$, as required by a PCP solution, holds iff $\<y|\gamma(u,w)|x\>=0$. If one wants to have a non-negative expression of this type (as in the following subsection), one can take the square, express it in terms of $\gamma\otimes\gamma$ and restrict to the anti-symmetric subspace---Lemma \ref{lem:PCPC} is based on such a representation (cf. \cite{BC03,Hir07} for more details).

One of the problems to which PCP can be reduced is the \emph{matrix mortality problem}. This asks whether a set of matrices $M:=\{M_i\in\cM_d(\mathbb{Z})\}_{i=1}^k$ admits a finite string of products for which $M_{i_1}\cdots M_{i_n}=0$. $M$ is then called 'mortal'. In \cite{EMG11} this has been applied to quantum information problems by constructing a set $M'\subset \cM_{d'}(\mathbb{Q})$ 
 which is mortal iff $M$ is, and which in addition can be interpreted as a set of 'Kraus operators' for which $\sum_j M_j'M_j'^\dagger=\1$. Note that by allowing algebraic entries, one can simply achieve this by setting $M_i':=X^{-1}M_iX/\sqrt{\lambda}$ using a positive eigenvector $\sum_i M_iX^2M_i^\dagger=\lambda X^2$ which is always guaranteed to exist.\footnote{If there is no positive definite eigenvector $X^2$, one can replace the $M_i$'s by a direct sum of their irreducible blocks, for which the corresponding positive map then has a positive definite eigenvector. The advantage of using $\R_{alg}$ instead of $\mathbb{Q}$ is that the number of matrices and their size doesn't change when going from $M$ to $M'$. In this way the argumentation in \cite{EMG11} then holds for  $(d,k)=(3,7)$ rather than only for $(15,9)$.}

\subparagraph{Diophantine equations.} A subset $D\in\mathbb{N}$ is called \emph{diophantine} iff for some $n\in\mathbb{N}$ there is a polynomial $p:\mathbb{N}^{n+1}\ra\mathbb{N}$ with integer coefficients, such that
\be
D=\{x|\exists y\in\mathbb{N}^n:p(x,y)=0\}.\ee
W.l.o.g. one can bound the degree of the polynomial by $4$ and, at the same time, the number of variables $n\leq 56$. A remarkable theorem \cite{DMR76}, which eventually proved the undecidability of Hilbert's 10th problem,  states that a set is diophantine if and only if it is recursively enumerable. Since membership is generally undecidable for the latter, the same has to hold for the former. This result should be compared with the earlier mentioned Tarski-Seidenberg theorem. While Thm.\ref{thm:TSQE} implies that the existence of roots of a system of polynomial equations with integer coefficients can be decided over $\R$, the same problem becomes undecidable over $\mathbb{N}$. Whether it is decidable over $\mathbb{Q}$ is still an open problem.\vspace*{5pt}

So much for general undecidability in a nutshell. Let's have a look at a specific problem and apply some of the mentioned results.

\subsection{Reachability of fidelity thresholds is undecidable}

As an example which naturally occurs in the context of quantum information theory, we will discuss the problem of deciding whether a small set of ``noisy gates'' allows to create a state which overcomes some pre-given fidelity threshold w.r.t. a given target state. The latter might for instance be a two-qubit singlet state and the aim, motivated by the requirements of entanglement distillation protocols, to achieve an overlap strictly larger than one half. 

In the following we will for simplicity let $\R$ and $\C$ denote the fields of real and complex algebraic numbers respectively. 

\begin{theorem}\label{thm:undecidable_fidelity} Let $(k,d)$ be a pair of integers which is either $(2,5)$ or $(5,3)$ (or pointwise larger than either) and take any real number $0<\lambda<1$ and normalized vector $|\phi\>\in\C^d$. Then, there is no algorithm which upon input of a density matrix $\rho\in\cM_d(\C)$ and of a set of quantum channels $\big\{T_i:\cM_d(\C)\ra\cM_d(\C)\big\}_{i=1}^k$ decides whether or not there is an integer $n\geq 1$ and a sequence $(i_1,\ldots,i_n)\in\{1,\ldots, k\}^n$ for which
\be \<\phi|\; T_{i_1}\cdots T_{i_n}(\rho)\;|\phi\> >\lambda. \label{eq:mainthmeq}\ee
\end{theorem}

The proof of this theorem is decomposed into two steps: (i) we establish a mapping from unconstrained matrices to quantum channels in such a way that concatenations of channels are closely linked to products of matrices, and (ii) we exploit known undecidability results for problems which are expressed in terms of products of matrices.

Let $\{H_i\in\cM_d(\C)\}_{i=1}^{d^2}$ be any Hermitian operator basis which is orthonormal in the sense that $\tr{H_i H_j}=\delta_{i,j}$ and satisfies  $H_1=\1/\sqrt{d}$. Note that the latter implies $\tr{H_i}=0$ for all $i>1$.

We represent any linear map  $T:\cM_d(\C)\ra\cM_d(\C)$ in terms of a $d^2\times d^2$ matrix with entries $\hat{T}_{i,j}:=\tr{H_iT(H_j)}$. $T$ is Hermiticity preserving iff $\hat{T}$ is real, and trace-preserving iff the first row of $\hat{T}$ is $(1,0,\ldots,0)$.

\begin{lemma}\label{lem:Tnuepsilon} Let $d\geq 2$ and $H_2:=(\1-d\psi)/(\sqrt{d^2-d})$ for any one-dimensional projector $\psi=|\psi\>\<\psi|\in\cM_d(\C)$. For every $M\in\cM_{d^2-2}(\R)$ and every $\nu\in(-\sqrt{d-1},1/\sqrt{d-1})$ there exists an $\epsilon>0$ such that
\be\hat{T}:=\left(\begin{array}{cc}
1 & 0 \\
\nu & 0
\end{array}\right)\oplus \epsilon M \label{eq:Tnuepsilon}\ee
represents a completely positive, trace-preserving linear map.\end{lemma}

 \proof{By construction $\hat{T}$ represents a trace-preserving and Hermiticity preserving linear map. First note that for $\epsilon=0$ the matrix $\hat{T}$ corresponds to the map
 \be T(X)=\frac{\tr{X}}{d}\left(\1+\frac{\nu}{\sqrt{d-1}}(\1-d\psi)\right).\ee
The Choi matrix of this map is positive definite iff
\be \1+\frac{\nu}{\sqrt{d-1}}(\1-d\psi) > 0, \ee
which holds in turn iff $\nu\in(-\sqrt{d-1},1/\sqrt{d-1})$. Within this range, full support of the Choi matrix then allows us to add a sufficiently small multiple of any other Hermiticity preserving map without violating complete positivity.
 }
\begin{proposition}\label{prop:MT} Let $d\geq 2$, $\lambda\in(0,1)$ and let $\phi=|\phi\>\<\phi|$ be a one-dimensional projector acting on  $\C^d$. For any $x,y\in\R^{d^2-2}$ and any set of matrices $\{M_i\in\cM_{d^2-2}(\R)\}_{i=1}^k$, one can construct a set of quantum channels $\{T_i:\cM_d(\C)\ra\cM_d(\C)\}_{i=1}^k$, a density matrix $\rho\in\cM_d(\C)$ and positive numbers $\epsilon,\delta>0$ such that for all natural numbers $n\geq 1$
\be \tr{\phi T_{i_1}\cdots T_{i_n}(\rho)}=\lambda+\delta\epsilon^n\<x|M_{i_1}\cdots M_{i_n}|y\>\ee
holds for all sequences $(i_1\ldots i_n)\in\{1,\ldots,k\}^n$.
\end{proposition}
\proof{First of all, we exploit the freedom in the choice of the Hermitian and orthonormal operator basis $\{H_i\}$. As indicated above, we fix $H_1:=\1/\sqrt{d}, H_2:=(\1-d\psi)/(\sqrt{d^2-d})$, with $\psi\neq\phi$ to be chosen later, and we choose $H_3,\ldots, H_{d^2}$ such that up to some  factor $\delta_1>0$  we have $\tr{\phi H_{i+2}}=\delta_1 x_i$ for all $i=1,\ldots d^2-2$.

We define $\rho\in\cM_d(\C)$ in this basis as being represented by the vector $\big(1/\sqrt{d},0,\delta_2y\big)$. In this way, $\rho=\rho^\dagger$, $\tr{\rho}=1$ and we can choose  $\delta_2>0$  sufficiently small, so that $\rho\geq 0$.

To every matrix $M_i$ we assign a quantum channel $T_i$ of the form in Eq.(\ref{eq:Tnuepsilon}) where a common $\epsilon>0$ is chosen such all the $T_i$'s are completely positive. In this way, we obtain
\bea \tr{\phi T_{i_1}\cdots T_{i_n}(\rho)} &=&\label{eq:lambdaterm} \frac{1}{d}\left(1+\nu\frac{1-d|\<\psi|\phi\>|^2}{\sqrt{d-1}}\right)\\
&& + \delta\epsilon^n\<x|M_{i_1}\cdots M_{i_n}|y\>,\eea
where we have set $\delta:=\delta_1\delta_2$. By choosing appropriate $\nu\in(-\sqrt{d-1},1/\sqrt{d-1})$ and $\psi\neq\phi$, the r.h.s. of Eq.(\ref{eq:lambdaterm}) can be set to any value in the open interval $(0,1)$ and Lemma \ref{lem:Tnuepsilon} guarantees that the $T_i$'s are completely positive and trace-preserving as requested.
}

The following Lemma is distilled from \cite{Hir07} where it has been proven by reduction from PCP with Claus instances:
\begin{lemma}\label{lem:PCPC}
Let $(k,m)$ be either $(5,6)$ or $(2,21)$. There is no algorithm which upon input of vectors $x,y\in\mathbb{Z}^m$ and of matrices $\{M_i\in\cM_m(\mathbb{Z})\}_{i=1}^k$ decides whether or not there is an integer $n\geq 1$ and a sequence $(i_1,\ldots,i_n)\in\{1,\ldots, k\}^n$ so that
\be \<x|M_{i_1}\cdots M_{i_n}|y\> >0.\ee
\end{lemma}

Together with the previous proposition this Lemma now proves Thm.\ref{thm:undecidable_fidelity}. Inserting a different classical result \cite{BC03, Hir07} into Prop.\ref{prop:MT} we obtain an analogous result in which $>$ is replaced by $\geq$ in Eq.(\ref{eq:mainthmeq}). Variation on the theme of Prop.\ref{prop:MT} also allows to use targets other than the overlap with a given pure state such as  the expectation value of some observable or the overlap with a given quantum channel.

\ \vspace*{5pt}

\section{Discussion}

Loosely speaking: while semi-algebraic problems are decidable, word problems are typically not. We have discussed to what extent this fact separates quantum information problems into decidable and undecidable ones. Many of the interesting problems, however, remain undecided. 

In the direction of proving decidability of some of them, one approach might be to look into tools which extend Tarski-Seidenberg in various directions. In \cite{VMM94} it has for instance been shown that quantifier elimination is possible for the real field augmented by exponentiation and all restricted analytic functions. Needless to say, there are also limitations on the possibility of quantifier elimination, which are discussed for instance in \cite{Gab07}. Also tools for deciding specific second-order theories might be worthwhile looking at \cite{Rab69}.

In the other direction, most of the known undecidable problems are, in one way or the other, word problems. It would therefore be interesting to see a word problem structure identified in one of the undecided problems we have discussed. Is there an undecidable word problem where strings are formed by tensor products rather than by ordinary products?\vspace*{8pt}

\emph{Acknowledgements:} We acknowledge financial support from the European projects COQUIT and QUEVADIS, the CHIST-ERA/BMBF project CQC, the Alfried Krupp von Bohlen und Halbach-Stiftung and the Spanish grants  I-MATH, MTM2008-01366, S2009/ESP-159 and QUITEMAD.

\bibliographystyle{alpha}

\begin{thebibliography}{JNP{\etalchar{+}}11}

\bibitem[Bar77]{Bar77}
Jon Barwise, editor.
\newblock {\em Handbook of mathematical logic}.
\newblock Elsevier, Amsterdam, 1977.

\bibitem[Bar02]{Bar02}
Jonathan Barrett.
\newblock Nonsequential positive-operator-valued measurements on entangled
  mixed states do not always violate a bell inequality.
\newblock {\em Phys. Rev. A}, 65:042302, 2002.

\bibitem[BC01]{BC03}
Vincent Blondel and Vincent Canterini.
\newblock Undecidable problems for probabilistic automata of fixed dimension.
\newblock {\em Theory of Computing Systems}, 36:231--245, 2001.

\bibitem[BCSS97]{BCSS97}
Leonore Blum, Felipe Cucker, Michael Shub, and Steve Smale.
\newblock {\em Complexity and real computation}.
\newblock Springer, 1997.

\bibitem[BJKP05]{BJKP05}
Vincent~D. Blondel, Emmanuel Jeandel, Pascal Koiran, and Natacha Portier.
\newblock Decidable and undecidable problems about quantum automata.
\newblock {\em SIAM J. Comput.}, 34:1464--1473, June 2005.

\bibitem[BPR94]{BPR94}
S.~Basu, R.~Pollack, and M.-F. Roy.
\newblock On the combinatorial and algebraic complexity of quantifier
  elimination.
\newblock {\em Foundations of Computer Science, 1994 Proceedings.}, pages 632
  -- 641, 1994.

\bibitem[CCH]{CCH09}
Toby~S. Cubitt, Jianxin Chen, and Aram~W. Harrow.
\newblock Superactivation of the asymptotic zero-error classical capacity of a
  quantum channel.
\newblock ar{X}iv:0906.2547, to appear in IEEE Trans.\ Inform.\ Theory.

\bibitem[CPGW]{CPW12}
Toby~S. Cubitt, David Perez-Garcia, and Michael~M. Wolf.
\newblock to be published.

\bibitem[Cut80]{Cut80}
Nigel~J. Cutland.
\newblock {\em Computability, an introduction to recursive function theory}.
\newblock Cambridge University Press, Cambridge, 1980.

\bibitem[DJK05]{DJK05}
Harm Derksen, Emmanuel Jeandel, and Pascal Koiran.
\newblock Quantum automata and algebraic groups.
\newblock {\em Journal of Symbolic Computation}, 39(3-4):357 -- 371, 2005.

\bibitem[DMM94]{VMM94}
Lou van~den Dries, Angus Macintyre, and David Marker.
\newblock The elementary theory of restricted analytic fields with
  exponentiation.
\newblock {\em The Annals of Mathematics}, 140(1):pp. 183--205, 1994.

\bibitem[DMR76]{DMR76}
M.~Davis, Yu. Matijacevic, and J.~Robinson.
\newblock Hilbert's tenth problem. Diophantine equations: positive aspects of a
  negative solution.
\newblock {\em Proc. Symposia Pure Math.}, 28:223--378, 1976.

\bibitem[DSW98]{DSW98}
Andreas Dolzmann, Thomas Sturm, and Volker Weispfenning.
\newblock Real quantifier elimination in practice.
\newblock In {\em Algorithmic Algebra and Number Theory}, pages 221--247.
  Springer, 1998.

\bibitem[Dua]{Run09}
Runyao Duan.
\newblock Super-activation of zero-error capacity of noisy quantum channels.
\newblock arXiv:0906.2527.

\bibitem[Fri]{Fri10}
Tobias Fritz.
\newblock Tsirelson's problem and Kirchberg's conjecture.
\newblock arXiv:1008.1168.

\bibitem[GAB07]{Gab07}
ANDREI GABRIELO.
\newblock Counterexamples to quantifier elimination for fewnomial and
  exponential expression.
\newblock {\em MOSCOW MATHEMATICAL JOURNAL}, 7:453--460, 2007.

\bibitem[Har09]{Har09}
Tero Harju.
\newblock Post correspondence problem and small dimensional matrices.
\newblock In Volker Diekert and Dirk Nowotka, editors, {\em Developments in
  Language Theory}, volume 5583 of {\em Lecture Notes in Computer Science},
  pages 39--46. Springer Berlin / Heidelberg, 2009.

\bibitem[Has09]{Has09}
M.~B. Hastings.
\newblock A counterexample to additivity of minimum output entropy.
\newblock {\em Nature Physics}, 5:255, 2009.
\newblock (arXiv:0809.3972 [quant\nobreakdash-ph]).

\bibitem[Hir07]{Hir07}
Mika Hirvensalo.
\newblock Improved undecidability results on the emptiness problem of
  probabilistic and quantum cut-point languages.
\newblock In Jan van Leeuwen, Giuseppe Italiano, Wiebe van~der Hoek, Christoph
  Meinel, Harald Sack, and Frantiöek Pl·öil, editors, {\em SOFSEM 2007: Theory
  and Practice of Computer Science}, volume 4362 of {\em Lecture Notes in
  Computer Science}, pages 309--319. Springer Berlin / Heidelberg, 2007.

\bibitem[Hir08]{Hir08}
Mika Hirvensalo.
\newblock Various aspects of finite quantum automata.
\newblock In Masami Ito and Masafumi Toyama, editors, {\em Developments in
  Language Theory}, volume 5257 of {\em Lecture Notes in Computer Science},
  pages 21--33. Springer Berlin / Heidelberg, 2008.

\bibitem[HM11]{HM11}
Uffe Haagerup and Magdalena Musat.
\newblock Factorization and dilation problems for completely positive maps on
  von neumann algebras.
\newblock {\em Communications in Mathematical Physics}, 303:555--594, 2011.

\bibitem[HW02]{HW02}
A.~S. Holevo and R.~F. Werner.
\newblock Counterexample to an additivity conjecture for output purity of
  quantum channels.
\newblock {\em J. Math. Phys.}, 43:4353--4357, 2002.
\newblock arXiv:quant-ph/0203003.

\bibitem[JE]{EMG11}
 J.~Eisert, M.P.~Mueller, C.~Gogolin.
\newblock Quantum measurement occurence is undecidable.
\newblock arXiv:1111.3965.

\bibitem[JNP{\etalchar{+}}11]{JNPPSW11}
M.~Junge, M.~Navascues, C.~Palazuelos, D.~Perez-Garcia, V.~B. Scholz, and R.~F.
  Werner.
\newblock Connes' embedding problem and Tsirelson's problem.
\newblock {\em Journal of Mathematical Physics}, 52(1):012102, 2011.

\bibitem[KO98]{KO98}
J.~K{\"o}rner and A.~Orlitsky.
\newblock Zero-error information theory.
\newblock {\em IEEE Trans. Inform. Theory}, 44(6):2207, 1998.

\bibitem[LM70]{LM70}
A.H. Lachlan and E.W. Madison.
\newblock Computable fields and arithmetically definable ordered fields.
\newblock {\em Proc. Amer. Math. Soc.}, 24:803--807, 1970.

\bibitem[Mad70]{Mad70}
E.~W. Madison.
\newblock A note on computable real fields.
\newblock {\em The Journal of Symbolic Logic}, 35(2):pp. 239--241, 1970.

\bibitem[Mar08]{Mar08}
Murray Marshall.
\newblock {\em Positive polynomials and sums of squares}.
\newblock AMS, 2008.

\bibitem[MS05]{MS05}
Yuri Matiyasevich and G\'{e}raud S\'{e}nizergues.
\newblock Decision problems for semi-Thue systems with a few rules.
\newblock {\em Theor. Comput. Sci.}, 330:145--169, January 2005.

\bibitem[MW09]{MW09}
Christian Mendl and Michael Wolf.
\newblock Unital quantum channels: convex structure and revivals of Birkhoff's
  theorem.
\newblock {\em Communications in Mathematical Physics}, 289:1057--1086, 2009.

\bibitem[NPA07]{NPA07}
Miguel Navascu\'es, Stefano Pironio, and Antonio Ac\'in.
\newblock Bounding the set of quantum correlations.
\newblock {\em Phys. Rev. Lett.}, 98:010401, Jan 2007.

\bibitem[O.L09]{Lev09}
O.Levin.
\newblock {\em Computability Theory, Reverse Mathematics, and Ordered Fields}.
\newblock PhD thesis, University of Connecticut, Storrs, CT, 2009.

\bibitem[Pat70]{Pat70}
M.S. Paterson.
\newblock Unsolvability in $3\times 3$ matrices.
\newblock {\em Stud. Appl. Math.}, 49:105--107, 1970.

\bibitem[PV10]{PV10}
K\'aroly~F. P\'al and Tam\'as V\'ertesi.
\newblock Maximal violation of a bipartite three-setting, two-outcome Bell
  inequality using infinite-dimensional quantum systems.
\newblock {\em Phys. Rev. A}, 82:022116, Aug 2010.

\bibitem[Rab69]{Rab69}
Michael~O. Rabin.
\newblock Decidability of second-order theories and automata on infinite trees.
\newblock {\em Transactions of the American Mathematical Society}, 141:pp.
  1--35, 1969.

\bibitem[Sei54]{Sei54}
A.~Seidenberg.
\newblock A new decision method for elementary algebra.
\newblock {\em Annals of Mathematics}, 60:365--374, 1954.

\bibitem[SST01]{SST00}
P.~W. Shor, J.~A. Smolin, and B.~M. Terhal.
\newblock Nonadditivity of bipartite distillable entanglement follows from
  conjecture on bound entangled Werner states.
\newblock {\em Phys.\ Rev.\ Lett.}, 86:2681, 2001.

\bibitem[SY08]{SY08}
Graeme Smith and Jon Yard.
\newblock Quantum communication with zero-capacity channels.
\newblock {\em Science}, 321:1812--1815, 2008.

\bibitem[Tar51]{Tar51}
A.~Tarski.
\newblock {\em A decision method for elementary algebra and geometry}.
\newblock University of California Press, 1951.

\bibitem[VW02]{VW02b}
Karl Gerd~H. Vollbrecht and Michael~M. Wolf.
\newblock Activating distillation with an infinitesimal amount of bound
  entanglement.
\newblock {\em Phys. Rev. Lett.}, 88:247901, May 2002.

\bibitem[Wat04]{Wat04}
John Watrous.
\newblock Many copies may be required for entanglement distillation.
\newblock {\em Phys.\ Rev.\ Lett.}, 93:010502, 2004.

\bibitem[Wer89]{Wer89}
R.~F. Werner.
\newblock Quantum states with {E}instein-{P}odolsky-{R}osen correlations
  admitting a hidden-variable model.
\newblock {\em Phys. Rev. A}, 40:4277, 1989.

\bibitem[WH08]{WH08}
A.~J. Winter and P.~Hayden.
\newblock Counterexamples to the maximal p-norm multiplicativity conjecture for
  all $p > 1$.
\newblock {\em Commun. Math. Phys.}, 284(1):263, 2008.
\newblock (arXiv:0807.4753 [quant\nobreakdash-ph]).

\bibitem[WW01]{WW01b}
R.~F. Werner and M.~M. Wolf.
\newblock Bell inequalities and entanglement.
\newblock {\em J. Quant.\ Inf.\ Comp.}, 1(3):1, 2001.

\end{thebibliography}

\newcommand{\etalchar}[1]{$^{#1}$}

\end{document}